\documentclass[prl,reprint,aps,amsmath,amssymb,twocolumn,superscriptaddress,bibliography]{revtex4-1}
\usepackage{graphicx}
\usepackage{caption}
\pdfoutput=1
\usepackage{overpic}
\usepackage{dcolumn}
\usepackage{bm}
\usepackage{subfigure}
\usepackage{amsmath}
\usepackage{amsfonts}
\usepackage{color}
\usepackage{array} 
\usepackage{booktabs} %
\usepackage{rotating}
\usepackage[english]{babel}
\usepackage{amsthm,amsmath,amssymb}
\usepackage{mathrsfs}
\usepackage{lipsum}
\usepackage{dutchcal}
\usepackage{float}
\usepackage{changes}
\usepackage[hidelinks]{hyperref}
\hypersetup{
    colorlinks=true,
    linkcolor=blue,
    filecolor=magenta,
    urlcolor=blue,
    citecolor=magenta,
    }   

\renewcommand{\Re}{\operatorname{Re}}
\renewcommand{\Im}{\operatorname{Im}}

\begin{document}

\title{Unveiling Stable One-dimensional Magnetic Solitons in Magnetic Bilayers}
\author{Xin-Wei Jin}
\affiliation{School of Physics, Northwest University, Xi'an 710127, China}
\affiliation{Peng Huanwu Center for Fundamental Theory, Xi'an 710127, China}

\author{Zhan-Ying Yang}
\email{zyyang@nwu.edu.cn}
\affiliation{School of Physics, Northwest University, Xi'an 710127, China}
\affiliation{Peng Huanwu Center for Fundamental Theory, Xi'an 710127, China}

\author{Zhimin Liao}
\affiliation{School of Physics, Peking University, Beijing, 100871,China}

\author{Guangyin Jing}
\email{jing@nwu.edu.cn}
\affiliation{School of Physics, Northwest University, Xi'an 710127, China}

\author{Wen-Li Yang}
\affiliation{Peng Huanwu Center for Fundamental Theory, Xi'an 710127, China}
\affiliation{Insititute of Physics, Northwest University, Xi'an 710127, China}

\date{\today}

\begin{abstract}
We propose a novel model which efficiently describes the magnetization dynamics in a magnetic bilayer system. By applying a particular gauge transformation to the Landau-Lifshitz-Gilbert (LLG) equation, we successfully convert the model into an exactly integrable framework. Thus the obtained analytical solutions allows us to predict a 1D magnetic soliton pair existed by tunning the thickness of the spacing layer between the two ferrimagnetic layers. The decoupling-unlocking-locking transition of soliton motion is determined at various interaction intensitiy. Our results have implications for the manipulation of magnetic solitons and the design of magnetic soliton-based logic devices.

\end{abstract}

\maketitle
\textit{Introduction.\textemdash}
The intricate interplay of multiple interactions in magnetic materials generates a large class of localized spin textures \textemdash magnetic solitons \cite{ahlberg2022freezing,moutafis2009dynamics,wang2023long,liu2018long,li2021topological,lan2021geometric,korber2020nonlocal,pribiag2007magnetic,ohkuma2020soliton,zhang2020skyrmion}. These solitons exhibit distinct and varied configurations in different dimensions and hold great promise as candidates for the next generation of magnetic storage devices \cite{gu2022three,zhang2017stateful}.
Instead of static magnetic interactions, dynamic magnetic interactions \cite{tserkovnyak2005nonlocal,klingler2018spin,gallardo2019reconfigurable} have been recently predict and observed by the current-induced torque or non-equilibrium spin pumping \cite{slonczewski1996current,apalkov2013spin,li2004domain,li2004domain2,liu2020three,heinrich2003dynamic,li2020coherent}. Within the dynamic coupling magnetic interaction, two magnets can be coherently and tunable coupled at the macro distance, presenting a novel avenue for the coherent transfer of magnon excitation between distinct magnetic systems \cite{li2020coherent,zhou2020ultrafast}.
Furthermore, these developments raises also an intriguing question of the existence and regulation of attractive magnetic solitons in magnetic bilayer structures \cite{yazdi2021tuning,zhang2016magnetic,xu2022systematic}.

Extensive efforts have been dedicated to the quest for stable magnetic solitons in theory, experiments, and micromagnetic simulations \cite{nadj2014observation,cai2023superconductor,sheng2023nonlocal,linder2015superconducting,tan2019propagation, liu2019current,yuan2022quantum,wang2022ferroelectric,shen2023programmable,yang2022spintronic,wang2023spintronic}.
The dynamics of magnetic solitons are described by the Landau-Lifshitz-Gilbert (LLG) equation \cite{heinrich2003dynamic,gilbert2004phenomenological}.
However, for decades, due to the intricate nature of this highly nonlinear coupled equations with multiple interactions, finding analytical solutions are extremely challenging and is a long-standing problem \cite{iacocca2017breaking,chen2022skyrmion}. The lack of comprehensive analytical solutions hinders progress, necessitating time-consuming and labor-intensive experiments and simulations, without the guidance of a solid theoretical framework.
The dynamic coupling magnetic interaction not only unveils a host of fresh physical phenomena but also amplifies the complexity of solving the coupled LLG equation from a theoretical standpoint.

In this letter, we establish an exchange-coupled magnetic bilayer structure, ferromagnetic/normal/ferromagnetic (F/N/F), as a model system.
From the coupled LLG equations governing the magnetization dynamics in the ferromagnetic bilayers, a theoretical model at small amplitude approximation is developed.
A gauge transformation is proposed allowing us to convert the problem into an integrable model, which is applicable when the intermediate layer thickness is appropriately chosen.
Thereafter, the exact solution of the governing equation is achineved, and the analytical magnetic soliton solutions are subsequently obtained.
By adjusting the strength of dynamic magnetic coupling, we find that the magnetic soliton pairs in the ferromagnetic bilayer undergo a decoupled-unlocking-locking transition.
We also examine the influence of Gilbert damping in materials on the design of practical devices.
These results illustrate practical ways to control the one-dimentional magnetic solitons, in which three motion states are successfully released: anti-parallel moving, splitting oscillation, and the locking soliton pair.

\begin{figure}[t]
\vspace{0cm} %
\includegraphics[width=0.45\textwidth]{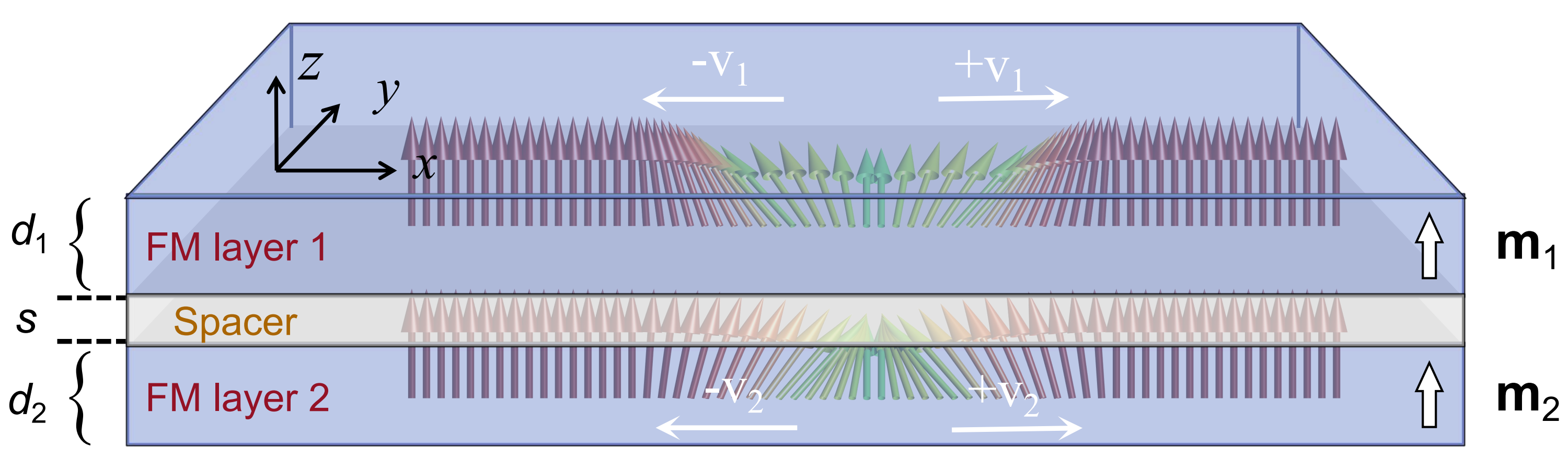}
\caption{Sketch of the ferromagnetic/normal/ferromagnetic thin film bilayer system. The magnetic soliton excitations propagate along $x$-axis. As a reference, the top (bottom) FM layer is labeled $i$ = 1($i$ = 2). Their corresponding thicknesses are represented by $d_{1}$ and $d_{2}$, respectively. Parameter $s$ denotes the thickness of the nonmagnetic interlayer.}\label{Sketch}
\end{figure}

\textit{Modeling.\textemdash}
We consider a magnetic bilayers system as illustrated in Fig. \ref{Sketch}, which consists of two coupled ferromagnetic (FM) films and a nonmagnetic interlayer with thicknesse of $s$.
The FM layers are assumed to be parallel to each other with equal thicknesses $d_{1}=d_{2}=d$.
The dynamics of the unit magnetization vector $\textbf{m}_{i}$ in the parallel coupled ferromagnetic layers can be described by the Landau-Lifshitz-Gilbert equation
\begin{equation}\label{LLG}
\frac{\partial\textbf{m}_{i}}{\partial t}=-\gamma_{i}\textbf{m}_{i}\times\textbf{H}^{i}_{\rm{eff}}+\alpha_{i}\left(\textbf{m}_{i}\times\frac{\partial\textbf{m}_{i}}{\partial t}\right)-\frac{\gamma_{i}J}{s M_{s,i}}\textbf{m}_{i}\times\textbf{m}_{j}.
\end{equation}
where $\gamma_{i}$ is the gyromagnetic ratio,
$\alpha_{i}>0$ denotes Gilbert damping parameter of each FM layer,
$M_{s,i}$ is the saturation magnetization,
and $J$ represents coupling strength between $\textbf{m}_{i}$ and $\textbf{m}_{j}$ with $i,j=1,2$.
Moreover, the effective field of the two FM layers can be obtained from the free energy density of the system as $\textbf{H}^{i}_{\rm{eff}}=-\frac{1}{\mu_{0}}\frac{\delta E}{\delta\textbf{m}_{i}}$.
%
We assume the total energy incorporates the contributions from the Zeeman energy due to an applied magnetic field $\textbf{H}_{0}=(0,0,h)$, the exchange interaction parametrized by an exchange constant $A_{i}$, and the perpendicular magnetic anisotropy energy.
Thus, it takes the form $\textbf{H}^{i}_{\rm{eff}}=\textbf{H}_{0}+(2A_{i}/M_{s,i})\nabla^{2}\textbf{m}_{i}+(2K_{i}/M_{s,i})(\textbf{m}_{i}\cdot\textbf{n})\textbf{n}$,
where $\textbf{n}=(0,0,1)$ is the unit vector directed along the anisotropy axis.
For simplicity, we transform the coupled LLG equation (\ref{LLG}) to the dimensionless form
$
\frac{\partial\textbf{m}_{i}}{\partial \tau}=-\textbf{m}_{i}\times\frac{\partial^{2}}{\partial \zeta^{2}}\textbf{m}_{i}-\kappa\textbf{m}_{i}\times(\textbf{m}_{i}\cdot\textbf{n})\textbf{n}-J'\textbf{m}_{i}\times\textbf{m}_{j},
$
by rescaling the space and time into $\zeta=\lambda^{-1}_{ex}\cdot x$, $\tau=\gamma\mu_{0}M_{s}\cdot t$. Here $\lambda_{ex}=\sqrt{2A_{i}/(\mu_{0}M_{s,i}^{2})}$ is the exchange length, $\kappa=2K_{i}/(\mu_{0}M_{s,i}^{2})$ and $J'=J/(\mu_{0}sM_{s,i}^{2})$ denote the dimensionless easy-plane anisotropy constant and dimensionless coupling strength, respectively.
Table \ref{Table} summarizes the realistic physical constants and parameters used for the structure under our consideration.

\begin{figure*}[ht]
\vspace{0cm} %
\centering
\includegraphics[width=17cm]{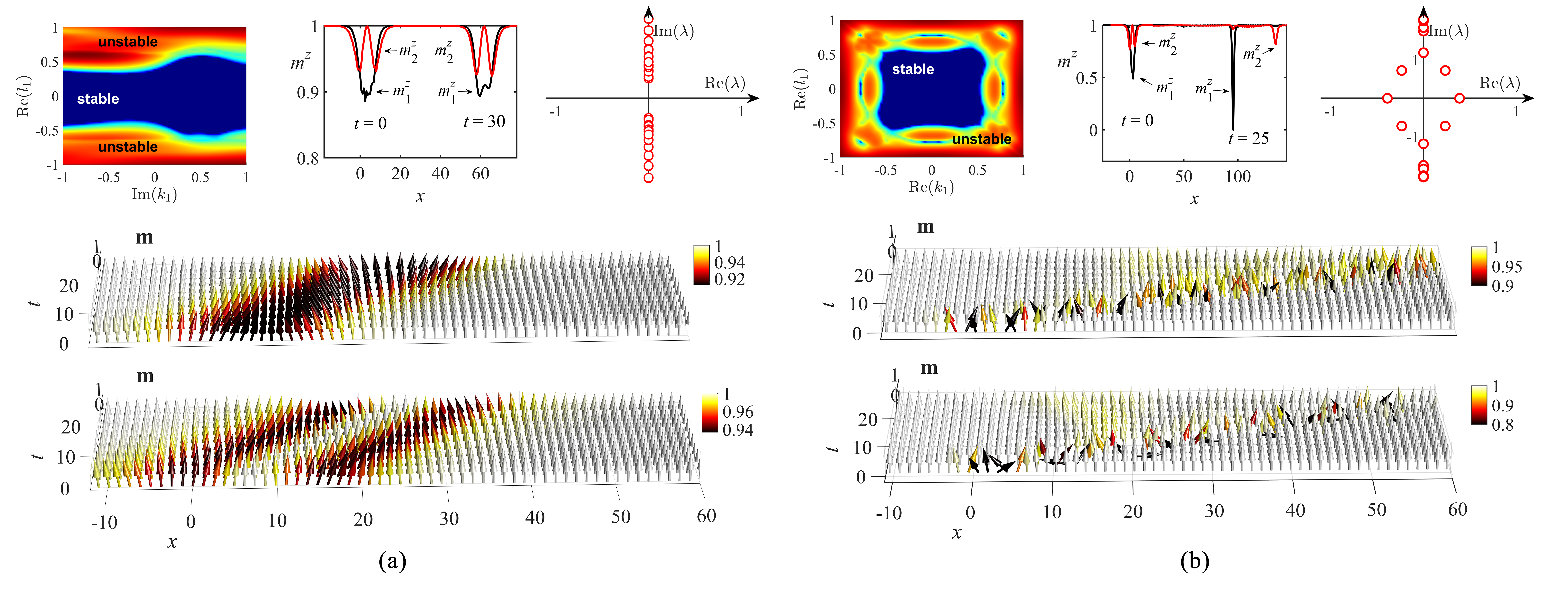}
\caption{Propagations of stable and unstable non-degenerate magnetic solitons. (a) Left panel: Stability regions in the parameter space $(\Im(k_{1}),\Re(l_{1}))$. Center panel: $m^{z}$ profiles of symmetric flat-bottom-double-hump magnetic soliton at $t=0$ and $t=30$. Right panel: eigenvalue spectrum. Bottom panel: stable propagations of magnetic soliton in two FM layers. (b) Left panel: Stability regions in the parameter space $(\Re(k_{1}),\Re(l_{1}))$. Center panel: $m^{z}$ profiles of asymmetric double-hump-double-hump magnetic soliton at $t=0$ and $t=30$. Right panel: eigenvalue spectrum. Bottom panel: unstable propagations of magnetic soliton in two FM layers.}\label{pro}
\end{figure*}

\begin{table}[b]
    \caption{The physical constants and parameters used.}
    \label{Table}
    \begin{ruledtabular}
    \resizebox{\linewidth}{!}{
    \begin{tabular}{lccc}
        Physical constants/parameters & Symbol & Value & Unit \\
        \colrule
        Gyromagnetic ratio & $\gamma$ & $1.76\times10^{11}$ & ${\rm \frac{rad}{s\cdot T}}$\\
        Saturation magnetization & $M_{s}$ & $5.8\times10^{5}$ & ${\rm \frac{A}{m}}$ \\
        Exchange stiffness & $A$ & $1.3\times10^{-11}$ & ${\rm \frac{J}{m}}$ \\
        Magnetic anisotropy & $K$ & $5\times10^{5}$ & ${\rm \frac{J}{m^{3}}}$ \\
        Magnetic permeability in vacuum & $\mu_{0}$ & $4\pi\times10^{-7}$ & ${\rm \frac{H}{m}}$ \\
        Damping parameter & $\alpha$ & $0.01 \sim 0.05$ &  \\
    \end{tabular}
    }
    \end{ruledtabular}
\end{table}

Take into account the fact that the magnitude of the magnetization $\textbf{m}_{i}^{2}=1$ at temperature well below the Curie temperature, we reasonably introduce a stereographic transformation
$\Phi_{j}=m_{j}^{x}+im_{j}^{y}, \left(m_{j}^{z}\right)^{2}=1-\lvert\Phi_{j}\rvert^{2}$.
Furthermore, let us consider small deviations of magnetization $\textbf{m}_{i}$ from the equilibrium direction (along the anisotropy axis), which corresponds to $\left(m_{j}^{x}\right)^{2}+\left(m_{j}^{y}\right)^{2}\ll \left(m_{j}^{z}\right)^{2}$ (or $\lvert\Phi_{j}\rvert^{2}\ll 1$) and therefore $m_{j}^{z}\approx1-\vert\Phi_{j}\vert^{2}/2$. As a result, the dynamics of the spinor ${\bf \Phi}=(\Phi_{1}, \Phi_{2})^{T}$ can be expressed as
\begin{equation}\label{mPhi}
\begin{split}
i\frac{\partial}{\partial \tau}{\bf \Phi}=\frac{\partial^{2}}{\partial \zeta^{2}}{\bf \Phi}+\left(J'\sigma_{1}-\Delta\right){\bf \Phi}+S\odot({\bf \Phi}{\bf \Phi}^{\dag}){\bf \Phi},
\end{split}
\end{equation}
where we have defined $S=\kappa (\sigma_{3})^{2}/2+J'\sigma_{1}/2$ and $\Delta=J'+h+2\kappa$, with $\sigma_{1,2,3}$ of the Pauli matrices.
Symbol $\odot$ represents the Hadamard product for matrices.
A noteworthy remark extracted here is that by maintaining a suitable separation between two ferromagnetic layers ($s=J/2K)$), it becomes possible to introduce a gauge transformation
$
\operatorname{\Phi_{1,2}}=\frac{1}{\sqrt{2}}\left(\Psi_{1}e^{i(h+\kappa)\tau}\pm\Psi_{2}e^{i(h+3\kappa)\tau}\right),
$
making the dynamic model (\ref{LLG}) entirely integrable.
Then, the new spinor ${\bf\Psi}=(\Psi_{1}, \Psi_{2})^{T}$ is determined by the Manakov equation with arbitrary constant coefficients:
\begin{equation}\label{manakov}
\begin{split}
i{\bf \Psi}_{\tau}={\bf \Psi}_{\zeta\zeta}+\kappa({\bf \Psi}{\bf \Psi}^{\dag}){\bf \Psi}.
\end{split}
\end{equation}
A diversity of solutions of this equation can be constructed using the methods of exactly integrable systems.
One can also easily obtain the formulations of three components of magnetization by the inverse transformation from Eq. (\ref{mPhi}).

\textit{Magnetic soliton solutions.\textemdash}Non-degenerate soliton solutions of (\ref{manakov}) can be constructed with the help of the Hirota bilinear formalism \cite{stalin2019nondegenerate, ramakrishnan2020nondegenerate}, and the first- and second- order non-degenerate soliton solutions are presented in \textit{Supplementary Materials}.
The final fundamental nondegenerate soliton solutions are characterized by four arbitrary complex parameters, describing the velocity and the amplitude of the magnetic soliton in both FM layers, as well as the nonlinear interaction of magnetic solitons between two FM layers.

From the non-degenerate soliton solution of Eq. (\ref{manakov}), the formulations of non-degenerate magnetic soliton are constructed.
This derived solutions represent several categories of magnetic solitons in this magnetic bilayer system.
Through analyzing these solutions, it becomes apparent that the magnetic bilayer system possesses diverse spin textures, manifested as dynamical magnetic solitons.
As far as we know, experimental observation of these magnetic soliton pairs resulting from interlayer dynamic interactions are currently lacking.
With this theoretical prediction, in the following, we try to discuss the possible generation mechanisms and the practical applications  by these magnetic soliton pairs in magnetic bilayer structures.

\textit{Linear stability analysis.\textemdash}It has been confirmed that, from the analytical solution above, there are magnetic solitons allowed in this system,  then another important aspect to be considered is their stability characters.
For practical applications of magnetic solitons as memory units or drivien objects in spintronics, it is crucial to maintain stability of solitons in the presence of interference.
The stability property is usually analyzed by way of linear stability analysis \cite{yang2010nonlinear, saha2020scalar, chen2021two}. For this purpose, we consider the solitary wave solutions of the form
${\bf \Psi}={\bf \Psi'}\exp(ib\tau)$,
with $b$ being propagation constant, then Eq. (5) becomes
\begin{equation}\label{Bmanakov}
-b{\bf \Psi'}_{\tau}={\bf \Psi'}_{\zeta\zeta}+\kappa({\bf \Psi'}{\bf \Psi'}^{\dag}){\bf \Psi'}.
\end{equation}
To analyze the linear stability of the solitary wave,
we perturb the relevant wave function as $\Psi_{i}=\left\{\Psi'_{0i}+[v_{i}(\zeta)+w_{i}(\zeta)]e^{\lambda \tau}+[v_{i}^{*}(\zeta)-w_{i}^{*}(\zeta)]e^{\lambda^{*} \tau}\right\}e^{ib\tau}$,
here $\Psi'_{0i}$ being the general complex-valued unperturbed wave function calculated from Eq. (\ref{manakov}), $v_{i}$ and $w_{i}(i=1,2)$ are small perturbations for a given eigenvalue $\lambda$.
Inserting this perturbed solution in Eq. (\ref{manakov}) and linearizing thereafter, we obtain the following linear-stability eigenvalue problem:
\begin{equation}
i {\textbf L}\cdot {\bf W}=\lambda \cdot {\bf W}.
\end{equation}
where matrix ${\bf W}=\left(v_{1},w_{1},v_{2},w_{2}\right)^{\rm T}$ denotes the normal-mode perturbations.
The matrix ${\textbf L}$ contains the magnetic soliton solution $\Psi'_{0i}$ representing the linear stability operator.
The matrix elements and calculation details of matrix {\textbf L} are presented in  \textit{Supplementary materials}.

In general, two separate regions can be defined based on the linear-stability spectrum. The non-degenerate soliton wave is linearly unstable when the spectrum contains eigenvalues with positive real parts, which gives an exponential growth rate of perturbations. While the soliton is regarded as stable if the spectrum contains purely imaginary discrete eigenvalues \cite{yang2010nonlinear}.
The whole spectrum of the linear-stability operator ${\textbf L}$ are numerically solved by the Fourier collocation method.

To verify the predictions of the linear stability analysis obtained from the numerical solution of the spectral problem (\ref{Bmanakov}), we
proceed to numerically simulate the nonlinear propagation of the magnetic solitons.
The evolutions of stable non-degenerate magnetic solitons and unstable non-degenerate magnetic solitons are illustrated in Figs. \ref{pro}.
The initial conditions for both simulations are taken in the form of a soliton solution perturbed by a $10\%$ random noise.
The upper panels of Fig. \ref{pro}(a) depict the stability regions in the parameter space $(\Im(k_{1}),\Re(l_{1}))$ of the magnetic soliton and provide an exemplary illustration of a stable soliton solution.
The center panel plots the shape of $m^{z}$ component in two ferromagnetic layers at $t=0$ and $t=30$. The whole stability spectrum of this non-degenerate soliton is shown in the upper right corner panel.
It can be seen that this flat-bottom-double-hump magnetic solitons propagate stably and the flat bottom structure in the first FM layer is maintained, which complies with the results
of the linear stability analysis.
On the other hand, Fig. \ref{pro}(b) shows the unstable propagation of the asymmetric single-double-hump soliton.
Stronger instabilities cause the splitting and diffusion of the solitons at relatively short times.

\begin{figure}[t]
\vspace{0cm} %
\centering
\includegraphics[width=8.5cm]{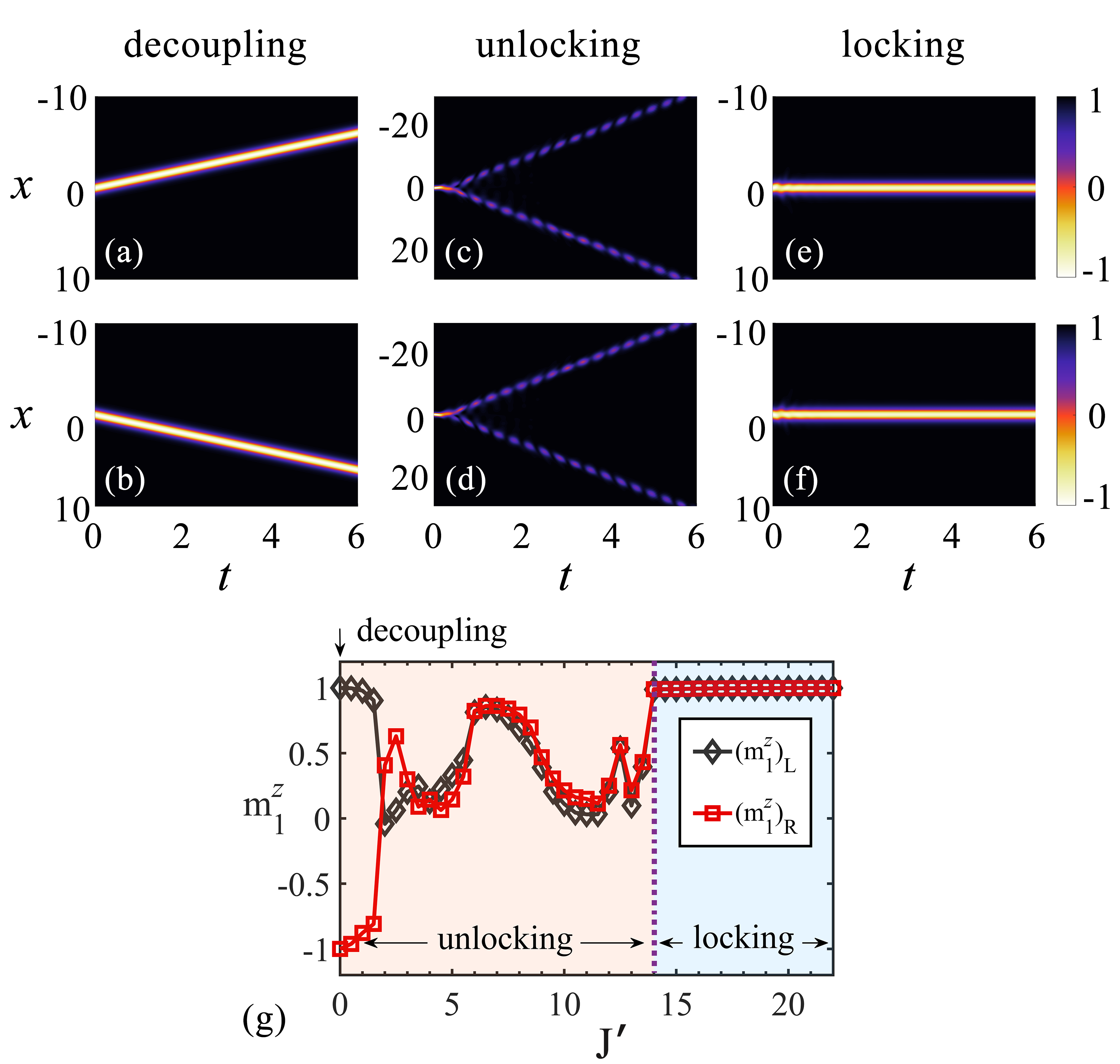}
\caption{The decoupling, unlocking and locking regions of magnetic soliton motion. (a)(b) Propagations of $m_{1}^{z}$ and $m_{2}^{z}$ with dimensionless coupling strength $J'=0$. (c)(d) Propagations of $m_{1}^{z}$ and $m_{2}^{z}$ with dimensionless coupling strength $J'=10$. (e)(f) Propagations of $m_{1}^{z}$ and $m_{2}^{z}$ with dimensionless coupling strength $J'=15$. (g) Phase diagram for the tristate transition by adjusting the interlayer coupling strength.}\label{transition}
\end{figure}

\textit{Coupling and Gilbert-damping.\textemdash} The successful stabilization of non-degenerate solitons enlightens us to design bilayer ferromagnetic spin-electronic devices based on stable magnetic solitons.
Here, we numerically investigate the propagation behavior of stable magnetic solitons in FM bilayers with various coupling strengths (which corresponds to thickness of the nonmagnetic spacer).

\begin{figure}[bp]
\vspace{0cm} %
\centering
\includegraphics[width=8.5cm]{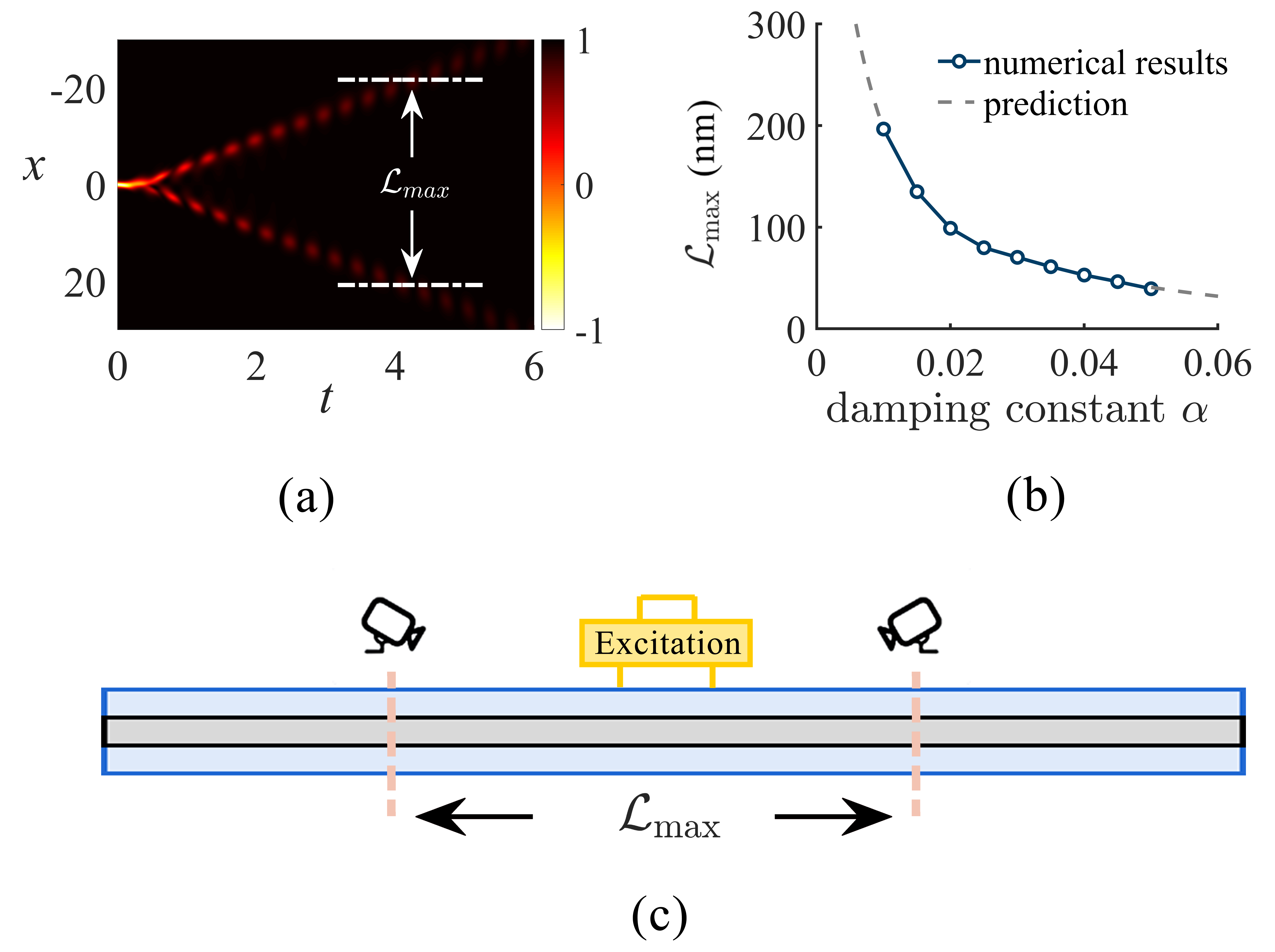}
\caption{The effect of Gilbert-damping on the motion of magnetic soliton in unlocking phase. (a) Propagations of $m_{1}^{z}$ with dimensionless coupling strength $J'=10$, damping constant $\alpha=0.05$, ${\mathcal L}_{\rm max}$ represents the maximum distance at which the signal attenuates to an unrecognizable state. (b) Dependence of the maximum distance ${\mathcal L}_{\rm max}$ for damping constant $\alpha$. (c) Sketch of maximum separation distance between identifiable magnetic soliton signals.}\label{Gilbert}
\end{figure}

Our first step is to construct a stable magnetic soliton in each layer, with opposing velocities.
When the two ferromagnetic layers are far apart from each other, their interaction becomes very weak, and the two layers are decoupling ($J'=0$). The two solitons propagate in opposite directions respectively, as depicted in Fig. \ref{transition}(a) and \ref{transition}(b). 
An increase of the coupling strength leads to soliton separation in both FM layers (as shown in Fig. \ref{transition}(c) and \ref{transition}(d)).
The interlayer interaction causes solitons to oscillate and propagate towards both ends at a constant velocity.
This observation can be explained as follows.
As the thickness of the intermediate layer reduces, the long-range dynamic interaction between the two FM layers, induced by adiabatic spin-pump, starts to come into play. The dynamic magnetization, which arises from the moving magnetic solitons in the ferromagnetic layer, causes the formation of non-equilibrium spin flow between the two layers. This ultimately triggers the bidirectional oscillation transmission of magnetic solitons.
We highlight that as the two ferromagnetic layers continue to approach, the interlayer dynamic interaction will exceed a certain threshold, which becomes sufficient to rapidly synchronize the motion of magnetic solitons and balance the spin current. Two solitons thereby get trapped in a stationary position (See Fig. \ref{transition}(e) and \ref{transition}(f)). This dynamic region of soliton immobilization is henceforth referred to as the locking region.
These simulation results in the wider range of $J'$ are summarized in Fig. \ref{transition}(g), which clearly shows the decoupling-to-unlocking-to-locking transition.
The black and red lines in the figure represent the minimum values of soliton signals received by the signal receiving devices placed at both ends of the first layer FM under different coupling strengths.

The different behaviors of magnetic solitons in FM bilayers under varying coupling strengths inspire us to design a logic signal generator. By adjusting the spacing between the two ferromagnetic layers, which is highly controllable in practice, it is possible to achieve different outputs of logical signals (Decoupling state corresponds to ``10" and ``01", unlocking state corresponds to ``11", and locking state corresponds to ``00"). This suggests a new posibility towards utilizing spintronic devices for logic operations. In practical applications, the signal attenuation caused by Gilbert damping in ferromagnetic materials must be considered. Through numerical simulation, we find that the damping effect has a significant impact on the magnetic solitons in the unlocking state. Fig. \ref{Gilbert}(a) shows the propagation of magnetic solitons in the unlocking state in the upper FM layer with Gilbert-damping constant $\alpha_{1}=0.05$, where ${\mathcal L}_{\rm max}$ represents the maximum distance at which the signal attenuates to an unrecognizable state (assuming that the $m^{z}$ component is greater than 0.8). The dependence of ${\mathcal L}_{\rm max}$ on the damping constant $\alpha$ for the FM layer is shown in Fig. \ref{Gilbert}(b). It can be observed that opting for materials featuring low damping coefficients can significantly increase the separation between signal receivers.

\textit{Discussion and Conclusion.\textemdash}To sum up, we have derived a model at a small amplitude approximation to describe the nonlinear dynamics of magnetization in a bilayer ferromagnetic system.
When the intermediate layer takes a characteristic thickness (i.e., 2 nm), $s=J/2K$ for the system here, and the dynamic interaction coupling parameter and magnetic anisotropy are taken as 2 {$\rm mJ/m^2$} and $5\times10^{5} {\rm J/m^3}$, it is possible to introduce a gauge transformation to transform the equation into a fully integrable constant coefficient Manakov system.
The first-order and second-order non-degenerate magnetic soliton solutions are obtained, as well as their respective stability regions. The numerical simulation results of magnetic soliton transmission are well consistent with the predictions given by linear stability.
These theoretical and numerical results confirm the existence of stable one-dimensional magnetic soliton pairs in magnetic bilayer system.
To generate such magnetic solitons in a F/N/F bilayer system, the magnetization texture based on the above magnetic soliton solution must be manufactured into the two ferromagnetic layers.
This excited solitions can be achieved for example by a local magnetic filed or spin-polarized electric currents.

On the other hand, the intensity of the interlayer long-range dynamic interaction, induced by adiabatic spin-pump, can be tailored by manipulating the spacing between the two FM layers.
Through the manipulation of the intermediate layer's thickness, we unveiled three distinct transport states of magnetic solitons: soliton decoupling, unlocking, and locking. With a gradual increment in dynamic interactions, we demonstrated the progression of magnetic soliton motion from decoupling to unlocking, and ultimately to locking.
It is note that the dynamic exchange coupling strength $J$ is related to the thickness of the spacing layer.
We postulate an inverse square root relationship between the two parameters \cite{li2020coherent}, i.e. $J\propto 1/\sqrt{s}$.
Through calculations based on the parameters we have considered, it is determined that when the thickness of the intermediate layer is less than 0.45 $\rm{nm}$, magnetic solitons initiate a transition towards the locking state.
Note that, the thickness of this transition is related to the selection of ferromagnetic layer and insulating spacer layer materials.
For the same material, various material properties such as the saturation magnetization $M_{s,i}$ and the interfacial dynamic coupling of the synthetic layers can be controlled within the reach of leading-edge material fabrication and deposition techniques \cite{zhang2016magnetic}.

Finally, we examine the impact of Gilbert damping in different ferromagnetic materials on this transitional process. Our findings reveal that damping predominantly results in the attenuation of magnetic solitons in the unlocking state.
Furthermore, we have established a correlation between the damping coefficient and the maximum separation distance between distinguishable magnetic soliton signals.
These findings present new possibilities for developing spintronic devices for logic computing based on magnetic solitons, and have ignited extensive research on these systems to refine their design according to specific application requirements.

\hspace*{\fill}

The authors thank Prof. H. M. Yu and  Prof. C. P. Liu for their helpful discussions. This work was supported by the National Natural Science Foundation of China (Nos. 12275213, 12174306, 12247103), and Natural Science Basic Research Program of Shaanxi (2023-JC-JQ-02, 2021JCW-19).

\bibliography{references}
\end{document}